\documentclass[journal]{IEEEtran}
\usepackage[pdftex]{graphicx}
\usepackage[cmex10]{amsmath}
\usepackage{amssymb}

\usepackage{caption}
\usepackage{subcaption}
\usepackage{cite}
\usepackage{psfrag}
\usepackage{epstopdf}
\usepackage{algorithm}
\usepackage{algorithmic}
\usepackage{widetext}
\usepackage{float}
\usepackage{color}
\usepackage{soul}
\usepackage{upgreek}
\usepackage{multirow}
\usepackage[justification=centerlast]{caption}
\usepackage{tikz,pgfplots}
\usepackage{dblfloatfix}
\usepackage{adjustbox}
\DeclareUnicodeCharacter{2005}{\hspace{0.01em}}
\usepackage{tikz} 
\usetikzlibrary{calc}
\usepackage{mathtools}
\usetikzlibrary{arrows}
\pgfplotsset{compat=newest}
\usepgfplotslibrary{fillbetween}
\usetikzlibrary{patterns}

\definecolor{myBlue}{RGB}{72,125,215}
\definecolor{myOrange}{RGB}{118,54,45}
\definecolor{InfinBlue}{RGB}{72,72,51}

\ifCLASSINFOpdf
\else
\fi
\begin{document}
%
% paper title
% can use linebreaks \\ within to get better formatting as desired
\title{Performance versus Complexity Study of Neural Network Equalizers in Coherent Optical Systems}
     \pgfplotsset{
        % use this `compat` level or higher to make use of the "advanced"
        % label positioning (this brings the second ylabel to the right)
        compat=1.3, 
        % (created a style for the common options)
        my axis style/.style={
            every axis plot post/.style={/pgf/number format/fixed},
            ybar=5pt,
            bar width=8pt,
            x=1.7cm,
            axis on top,
            enlarge x limits=0.1,
            symbolic x coords={MLP, biLSTM, ESN, CNN+MLP, CNN+biLSTM, DBP},
            %restrict y to domain*=0:1200, % Cut values off at 14
            visualization depends on=rawy\as\rawy, % Save the unclipped values
%            after end axis/.code={ % Draw line indicating break
%                \draw [ultra thick, white, decoration={snake, amplitude=1pt}, decorate] (rel axis cs:0,1.05) -- (rel axis cs:1,1.05);
%            },
            nodes near coords={%
                \pgfmathprintnumber[precision=2]{\rawy}% Print unclipped values
            },
            every node near coord/.append style={rotate=90, anchor=west},
            tick label style={font=\footnotesize},
            xtick distance=1,
        },
    }
% author names and IEEE memberships
% note positions of commas and nonbreaking spaces ( ~ ) LaTeX will not break
% a structure at a ~ so this keeps an author's name from being broken across
% two lines.
% use \thanks{} to gain access to the first footnote area
% a separate \thanks must be used for each paragraph as LaTeX2e's \thanks
% was not built to handle multiple paragraphs
%

\author{Pedro J. Freire, Yevhenii Osadchuk, Bernhard Spinnler, Antonio Napoli, Wolfgang Schairer, Nelson Costa, Jaroslaw E. Prilepsky, Sergei K. Turitsyn
\thanks{This paper was supported by the EU  Horizon 2020 program under the Marie Sklodowska-Curie grant agreement 813144 (REAL-NET). YO  acknowledges the support of the SMARTNET EMJMD programme (Project number - 586686-EPP-1-2017-1-UK-EPPKA1-JMD-MOB). JEP is supported by Leverhulme Trust, Grant No. RP-2018-063. SKT acknowledges support of the EPSRC project TRANSNET}
\thanks{Pedro J. Freire, Yevhenii Osadchuk, Jaroslaw E. Prilepsky  and Sergei K. Turitsyn are with Aston Institute of Photonic Technologies, Aston University, United Kingdom, p.freiredecarvalhosouza@aston.ac.uk.}
\thanks{Antonio Napoli, Wolfgang Schairer and  Bernhard Spinnler are with Infinera R\&D, Sankt-Martin-Str. 76, 81541, Munich, Germany, anapoli@infinera.com.}
\thanks{Nelson Costa is with Infinera Unipessoal, Lda, Rua da Garagem nº1, 2790-078 Carnaxide, Portugal, ncosta@infinera.com.}
\thanks{Manuscript received xxx 19, zzz; revised January 11, yyy.}}

% note the % following the last \IEEEmembership and also \thanks - 
% these prevent an unwanted space from occurring between the last author name
% and the end of the author line. i.e., if you had this:
% 
% \author{....lastname \thanks{...} \thanks{...} }
%                     ^------------^------------^----Do not want these spaces!
%
% a space would be appended to the last name and could cause every name on that
% line to be shifted left slightly. This is one of those "LaTeX things". For
% instance, "\textbf{A} \textbf{B}" will typeset as "A B" not "AB". To get
% "AB" then you have to do: "\textbf{A}\textbf{B}"
% \thanks is no different in this regard, so shield the last } of each \thanks
% that ends a line with a % and do not let a space in before the next \thanks.
% Spaces after \IEEEmembership other than the last one are OK (and needed) as
% you are supposed to have spaces between the names. For what it is worth,
% this is a minor point as most people would not even notice if the said evil
% space somehow managed to creep in.

% The paper headers
\markboth{Journal of Lightwave technology , ~Vol.~y, No.~x, February~2021}%
{Shell \MakeLowercase{\textit{et al.}}:  Computational Complexity Study of Neural Network Channel Equalizers in Optical Systems}
% The only time the second header will appear is for the odd numbered pages
% after the title page when using the twoside option.
% 
% *** Note that you probably will NOT want to include the author's ***
% *** name in the headers of peer review papers.                   ***
% You can use \ifCLASSOPTIONpeerreview for conditional compilation here if
% you desire.

% If you want to put a publisher's ID mark on the page you can do it like
% this:
%\IEEEpubid{0000--0000/00\$00.00~\copyright~2007 IEEE}
% Remember, if you use this you must call \IEEEpubidadjcol in the second
% column for its text to clear the IEEEpubid mark.

% use for special paper notices 
%\IEEEspecialpapernotice{(Invited Paper)}

% make the title area
\maketitle
\begin{abstract}
We present the results of the comparative performance-versus-complexity analysis for the several types of artificial neural networks (NNs) used for nonlinear channel equalization in coherent optical communication systems. The comparison is carried out using an experimental set-up with the transmission dominated by the Kerr nonlinearity and component imperfections.
For the first time, we investigate the application to the channel equalization of the convolution layer (CNN) in combination with a bidirectional long short-term memory (biLSTM) layer and the design combining CNN with a multi-layer perceptron. Their performance is compared with the one delivered by the previously proposed NN-based equalizers: one biLSTM layer, three-dense-layer perceptron, and the echo state network. Importantly, all architectures have been initially optimized by a Bayesian optimizer. First, we present the general expressions for the computational complexity associated with each NN type; these are given in terms of real multiplications per symbol. We demonstrate that in the experimental system considered, the convolutional layer coupled with the biLSTM (CNN+biLSTM) provides the largest Q-factor improvement compared to the reference linear chromatic dispersion compensation (2.9~dB improvement). Then, we examine the trade-off between the computational complexity and performance of all equalizers and demonstrate that the CNN+biLSTM is the best option when the computational complexity is not constrained, while when we restrict the complexity to some lower levels, the three-layer perceptron provides the best performance.  
\end{abstract}
% IEEEtran.cls defaults to using nonbold math in the Abstract.
% This preserves the distinction between vectors and scalars. However,
% if the journal you are submitting to favors bold math in the abstract,
% then you can use LaTeX's standard command \boldmath at the very start
% of the abstract to achieve this. Many IEEE journals frown on math
% in the abstract anyway.

% Note that keywords are not RMpSally used for peerreview papers.

% For peer review papers, you can put extra information on the cover
% page as needed:
% \ifCLASSOPTIONpeerreview
% \begin{center} \bfseries EDICS Category: 3-BBND \end{center}
% \fi
%
% For peerreview papers, this IEEEtran command inserts a page break an
% creates the second title. It will be ignored for other modes.
\IEEEpeerreviewmaketitle

\section{Introduction}

\IEEEPARstart{A}{mongst} the variety of different nonlinearity compensation methods, the machine learning (ML) based techniques are gaining momentum as a promising and flexible tool capable to efficiently unroll fiber and component-induced impairments. In the past several years, the research on artificial neural networks (NN) for optical channel equalization has already led to the development of a noticeable number of novel digital signal processing (DSP) methods that can provide the performance better than that rendered by the  ``conventional'' DSP approaches~\cite{6975096,WANG20171, hager2018nonlinear, 9083434, Bitachon20, 9184798, melek2020nonlinearity, zhang2019field, freire2020complex, schaedlerrecurrent}. The fast development of NN-related research and the growing ML developers community incites testing different novel NN architectures to mitigate fiber propagation impairments. In terms of the experimental verification of NN-based equalizers, several works dealt with the intensity-modulation with direct-detection (IM/DD) links. It was demonstrated that the application of the NNs with different internal structure, such as multi-layer perceptron (MLP)~\cite{estaran2016artificial, yi2020neural} (i.e. a simple densely connected feed-forward NN architecture), convolutional NNs (CNN)~\cite{Li:09, liang2020research}, echo state networks (ESN)~\cite{ranzini2020optoelectronic}, and long short-term memory (LSTM) NNs~\cite{dai2019lstm}, is  efficient in improving optical system-level performance. However, the test of similar NN architectures in coherent optical systems has been carried out, mainly, numerically~\cite{deligiannidis2020compensation, 9324921, sidelnikov2019methods, hager2020physics}, or in short-haul experiments~\cite{ranzini2021experimental, Schaedler2019deep, 9195215, bajaj2020single}. It is worth noticing that some very recent works evaluated the functioning of NN-based equalizers in metro/long-haul trials~\cite{zhang2019field, freire2020complex, 9083434, Bitachon20, schaedlerrecurrent}.

The variety of existing and emerging channel equalizers makes a comparative analysis of the different solutions a timely challenge. The NN-based channel equalization refers to two important aspects: i) the improvement of performance by the reduction of bit-error-rate (BER), and ii) the complexity of the algorithms, which is a fundamental issue for practical implementation. Clearly, the comparison can be carried out only for specific systems: some approaches can be more suitable for certain transmission links, while the others are favorable for different systems.

To gain a thorough understanding of how each of the aforementioned NN architectures performs, we need to pick a benchmark system for the comparison. In this work, we perform such a comparison using, as a benchmark, a single channel transmission of a dual-polarization (DP) 16-QAM signal with 34.4~GBd rate transmitted over 9$\times$50km TrueWave Classic (TWC) fiber spans at the power of 2 dBm. Such a choice of the fiber and the power level ensures that the system is in the strongly nonlinear regime, as we intend to study how the NNs unroll the Kerr nonlinearity effects. In our work, we analyze both the synthetically simulated and the experimental data. We, first, analyze the performance of several previously studied NN models: MLP, bidirectional LSTM (biLSTM), and ESN. Next, we compare their performance with that rendered by new composite NN structures: i) the convolutional layer coupled with the MLP (CNN+MLP); ii) the combination of the convolutional layer with the biLSTM (CNN+biLSTM). These new designs are, then, tested in the same environment, allowing us to infer the performance characteristics pertinent to each type among the 7 different NN topologies. We point out that the term ``topology'' in our research identifies \textit{the particular NN structure (architecture) with a specific fixed distribution of hyper-parameters}. We emphasize that, in contrast to other similar investigations, we employ \textit{the Bayesian optimization procedure} ~\cite{freire2020complex} for each NN type studied. This provides the optimal distribution of hyper-parameters pertaining to each NN type, such that we identify the best functioning regime (in terms of the performance delivered) for each architecture without complexity constraints. We show that the new CNN+biLSTM combination performs better than all other studied types. For each NN type considered hereafter, we also present the analytical expressions for the complexity, i.e. the number of multiplications attributed to each specific NN per recovered symbol. The highest complexity for the optimized NN equalizers corresponds to the new CNN+biLSTM composition that also renders the best performance. 

The completely new subject in the remit of this manuscript is what happens when we restrain the complexity of different NN types: no work has previously addressed the comparison of the performance rendered by different NNs considering \textit{the identical levels of computational complexity}. Our findings demonstrate a nontrivial behavior: while at the relatively high complexity levels the best performing model is the CNN+biLSTM, when we constraint the complexity to lower values, the simple MLP equalizer outperforms the advanced NN structures with the same complexity. Nevertheless, we notice that the goal of this paper is not to reach a broad conclusion about the trade-off between the complexity and  performance for all possible transmission scenarios; rather, we aim at emphasizing the importance of accounting for this issue in the equalizer design stage, and we provide the tools for one's correctly assessing the DSP-type complexity of the most popular neuro-layers.

The paper is organized as follows. In Sec.~\ref{sec:zoo} we describe the details of the different NN equalizers analyzed in our study. Sec.~\ref{Sec:complexity} presents how to compute the computational complexity on all NN-based equalizers considered in this paper. Sec.~\ref{Sec:results} describes the experimental setup and contains the results, including the comparison between the performance and computational complexity of different NN topologies; the performance is also compared with the digital back-propagation with 3 steps per span. Our findings are summarized in the conclusion.

\section{A zoo of neural network-based equalizers}\label{sec:zoo}
In this section, we revisit the most popular NN architectures that have been proposed and investigated so far in coherent optical channel post-equalization. We also introduce two new composite NN equalizer structures that can be deemed as the extension of previously proposed NN configurations.

To enhance the reproducibility of our methods, we provide a thorough summary of each NN architecture. The code of the algorithms implemented in Python 3.6.9 with TensorFlow (2.2.0) GPU backend and Keras (2.3.1), is provided in Zenodo~\cite{pedro_jorge_freire_de_carvalho_souza_2021_4582298}.

Before addressing the details of the NN-based equalizers, let us describe how the datasets used in this work are created. When dealing with the optical channel equalization, we require the NN to process not only the symbol of interest but also the neighboring ones insofar as both the chromatic dispersion and the drive amplifier add the memory to the channel. The latter means that the NN performs better if it is given information about the correlations between the symbols in the sequence. Therefore, the input of the real-value NN models used in this paper (in the regression task), is the time-domain vector delayed by $k$ symbols (the memory vector) containing the real and imaginary parts of both polarizations for the symbol at the time-step $k$ and its $2N$ neighboring (past and future) symbols. In the NN signal processing, due to the computational memory constraints the input layer receives just a portion of the total data, called the mini-batch, as far as the finite computational resources limit the length of the sequences with which we can operate. The NN input mini-batch shape can be defined by three dimensions: $(B, M, 4)$, where $B$ is the mini-batch size, $M$ is the memory size defined through the number of neighbors $N$ as $M = 2N + 1$, and $4$ is the number of features for each symbol, referring to the real and imaginary parts of two polarization components. The output target is to recover the real and imaginary parts of the $k$-th symbol of one of the polarization, so the shape of the NN output batch can be expressed as $(B,2)$.

In general, for all the NNs considered in this paper, we use the mean square error (MSE) loss estimator, since this choice corresponds to the conventional loss function frequently used for the regression tasks~\cite{aggarwal2018neural}. The other types of loss functions such as the mean absolute error, the Huber Loss, and the Log-Cost loss, were also considered for our NNs, but they did not show any noticeable benefits compared to the MSE. Moreover, it is important to highlight that we decided to present just the regression task in this paper because (for our test case scenario) the results achieved by regression and classification algorithms were close, but some fewer epochs were needed in the case of regression to reach the lowest BER. 

The classical Adam algorithm was chosen for the stochastic optimization step with the default learning rate equal to 0.001~\cite{gulli2017deep}. All NNs were trained for at most 1000 epochs (if not stopped earlier because of negligible changes in the loss function value over 150 epochs) and, after every training epoch, we calculated the BER obtained using the independently generated testing dataset.

The dataset was composed of $2^{20}$ symbols for the training dataset and $2^{18}$ independently generated symbols for the evaluation. To eliminate any possible data periodicity and overestimation~\cite{eriksson2017applying} in our experiment, a pseudo-random bit sequence (PRBS) of order 32 was used to generate those datasets with different random seeds for each of them. The periodicity of the data is, therefore, $2^{10}$ times higher than our training dataset size, since the modulation format used in our study was the 16 QAM. For the simulation, the Mersenne twister generator \cite{matsumoto1998mersenne}, that has periodicity equal to  $2^{19937} - 1$, was used with different random seed. Additionally, we highlight that the NN training data were shuffled using \texttt{numpy.random.shuffle} function in Python before feeding the dataset into the NN: such a shuffling helps to mitigate overfitting. The experimental setups and scenarios in which the datasets were acquired are described in the following sections. 

The following subsections will delve deeper into the design of the NN models used within this paper.

\subsection{A multi-layer perceptron}\label{subsec:MLP}
The first and, perhaps, simplest and well-studied NN-based equalizer that we consider is the MLP, proposed for the short-haul coherent system equalization in~\cite{Schaedler2019deep} and the long-haul systems in~\cite{sidelnikov2018equalization}. The MLP is a deep feed-forward densely connected NN structure that handles the I/Q components for each polarization jointly, providing two outputs for each processed symbol: its real and imaginary parts. Due to the MLP's ability to process joint I/Q components, the equalizer can learn the nonlinear phase impairments in addition to the amplitude-related nonlinearities. When using the MLP, the channel and device-induced memory effects are taken into account by incorporating the time-delayed versions of the input signal, as was carried out in~\cite{sidelnikov2018equalization}. 

In a simulation environment, the MLP equalizer showed performance metrics similar to those delivered by the ``traditional'' digital back-propagation (DBP) with 2 steps-per-span and 2 samples-per-symbol at 1000 km of standard single-mode fiber~\cite{sidelnikov2018equalization}. In our current paper, we use the same 3-layer MLP as in~\cite{Schaedler2019deep}, but in our case here \textit{the number of neurons and the activation function optimized for each layer}. Importantly, the number of layers in MLP, which is 3, has been found as optimal for our particular transmission scenario by the Bayesian optimizer (BO). However, this MLP topology rendered the BO, can alter essentially for different transmission scenarios. 

The general equation, in a matrix form, describing the output vector $y$ given the input $x$ passing through the 3-layer MLP, is:
\begin{equation}\label{eq:MLP}
 y\! =\! \phi \Bigg\{ \phi \Big[ \phi (x \! \times \! W_{n_1}\! + \! b_{1} ) \! \times\! W_{n_2}\! +\! b_{2} \Big] \!\times \!W_{n_3}\! +\! b_{3} \Bigg\}\! \times \! W_{out},
\end{equation}
where $x$ is the input vector with $n_i$ elements, $y$ is the output vector with $n_o$ elements, $\phi$ is a nonlinear activation function, $W_{n_1} \in \mathbb{R}^{n_i \times n_1}$, $W_{n_2} \in \mathbb{R}^{n_1 \times n_2}$, $W_{n_3} \in \mathbb{R}^{n_2 \times n_3}$ and $W_{out} \in \mathbb{R}^{n_3 \times n_o}$ are the real weight matrices of the respective dimensions participating in each layer of the MLP, $b_{1,2,3}$ are the bias vectors,  the indexes $n_{1,2,3}$ stand for the number of neurons in each hidden layer, and $\times$ in (\ref{eq:MLP}) is the matrix-vector convolution.

 \subsection{Long short-term memory NNs}\label{subsec:biLSTM}
 
\begin{figure}[thb!]
\centering\includegraphics[width=8.7cm]{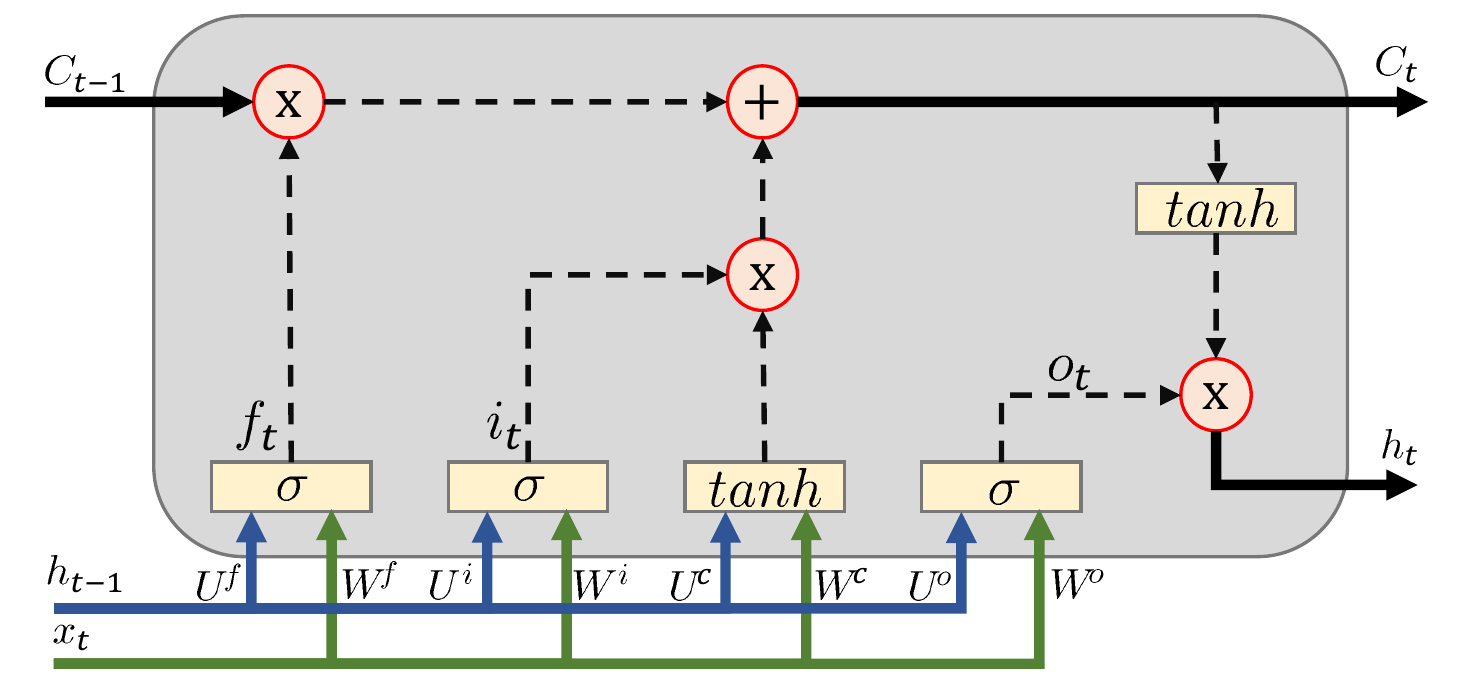}
\caption{The Long Short-Term Memory (LSTM) cell diagram representing graphically the operations described by (\ref{EQ.LSTM}) for one time-step. The arrows represent the ``flow'' of respective variables (the blue/green ones refer to the previous state and current input), the rectangles identify the nonlinear functions, while the symbols in circles identify the respective mathematical operations.}
\label{LSTMcell}
\end{figure}

 Compared to static (memoryless) systems where the MLPs can be efficient, the time sequences usually ought to be approached dynamically. Thus, recurrent NNs (RNNs) are often favored over other NN models for time sequences. However, training the recurring connections can be a much more complicated task compared to the MLP training, so that the network weights are usually changed almost imperceptibly. This aspect of RNNs often leads to the well-known vanishing gradient problem~\cite{Graves:11,aggarwal2018neural}. The LSTM networks were built to solve it and to harness the memory-related effects. The LSTM comprises a gateway architecture that includes three gate types: the input ($i_t$) gates, the forget ($f_t$) gates, and the output ($o_t$) gates, as shown in Fig.~\ref{LSTMcell}. The compact form of the forward pass LSTM cell equations for a time-step $t$, (i.e. when we process the input feature sequence \(x_{t}\) having the size $n_i$) is~\cite{karim2017lstm,greff2016lstm}: 
\begin{equation}\label{EQ.LSTM}
    \begin{gathered}
i_{t} = \sigma(W^{i}{x}_{t} + U^{i}{h}_{t-1} ),  \\
    f_{t} = \sigma(W^{f}{x}_{t} + U^{f}{h}_{t-1} ), \\
o_{t} = \sigma(W^{o}{x}_{t} + U^{o}{h}_{t-1} ),\\
    C_{t} = f_{t}\odot C_{t-1} + i_{t}\odot \tanh(W^{c}{x}_{t} + U^{c}{h}_{t-1}), \\
    h_{t} = o_{t} \odot \tanh(C_{t}),
    \end{gathered}
\end{equation}
where $C_t$ is the cell state vector, \(h_{t}\) is the current hidden state vector of the cell with size $n_h$ and $h_{t-1}$ is the previous hidden state vector. Note that $n_i$ is equal to the number of features, and $n_h$ is the number of hidden units that will be chosen in the design process. The trainable parameters of the LSTM network are represented by the matrices $W \in \mathbb{R}^{n_h \times n_i} $ and $U \in \mathbb{R}^{n_h \times n_h} $ with the respective upper indices $i$, $f$, $o$, and $c$, referring to the particular LSTM gates mentioned previously. More details are given in Fig.~\ref{LSTMcell}. In (\ref{EQ.LSTM}), \(\odot\) is the element-wise product, and \(\sigma\) denotes the logistic sigmoid activation function. The aim of the \textit{input} $i$-gate is to store the content to the cell; the \textit{forget} $f$-gate defines what information is to be erased; the \textit{output} $o$-gate defines what information has to be passed to the next cell. 

What makes the usage of the LSTM a dynamical approach is: the time sequence is processed by the array of LSTM cells ranging over the $t$-interval of interest, which is the memory size in our case. Besides the ``dynamical'' LSTM property, the \textit{bidirectional} LSTM (biLSTM) provides a more robust solution for time series since with the bidirectional structure, we are learning the weights from the past visible values to the future hidden values, and that corresponds to our learning which features of the past values are useful for a particular symbol value prediction~\cite{osogami2017bidirectional}. In the optical channel equalization context, the key advantage of biLSTM is that it can efficiently handle intersymbol interference (ISI) between the preceding and the following symbols.

In the context of channel equalization, the LSTM was suggested in~\cite{Xiaoxiao:07,lu2019memory} to reduce the transmission impairments in IM/DD systems with pulse-amplitude modulation (PAM). The LSTM-based approach was developed further in~\cite{deligiannidis2020compensation}, where, for the first time, the biLSTM was used in an optical coherent system to compensate for fiber nonlinearities, but only in a simulation environment. Additionally, it was shown that the biLSTM also outperformed a low-complexity DBP~\cite{deligiannidis2020compensation}. More recently, a bi-vanilla RNN  was applied as well for the soft-demapping nonlinear ISI~\cite{schaedlerrecurrent}.
In our current study, we use a similar structure as in~\cite{deligiannidis2020compensation}
, where the NN model is made up of a bidirectional LSTM layer followed by a dense layer. Finally, we note that, in contrast to the previous studies where the grid search was executed to guess the optimal number of hidden unities and memory size, this paper uses the BO to identify the best-performing biLSTM structure~\cite{freire2020complex}. 

\subsection{Echo state networks}
%  Recently, paper~\cite{grigoryeva2018echo} proved that the ESNs are universal function approximators, which potentially makes this approach as powerful as the other complicated NN structures.

\begin{figure}[ht!]
\centering
\includegraphics[width=8.5cm]{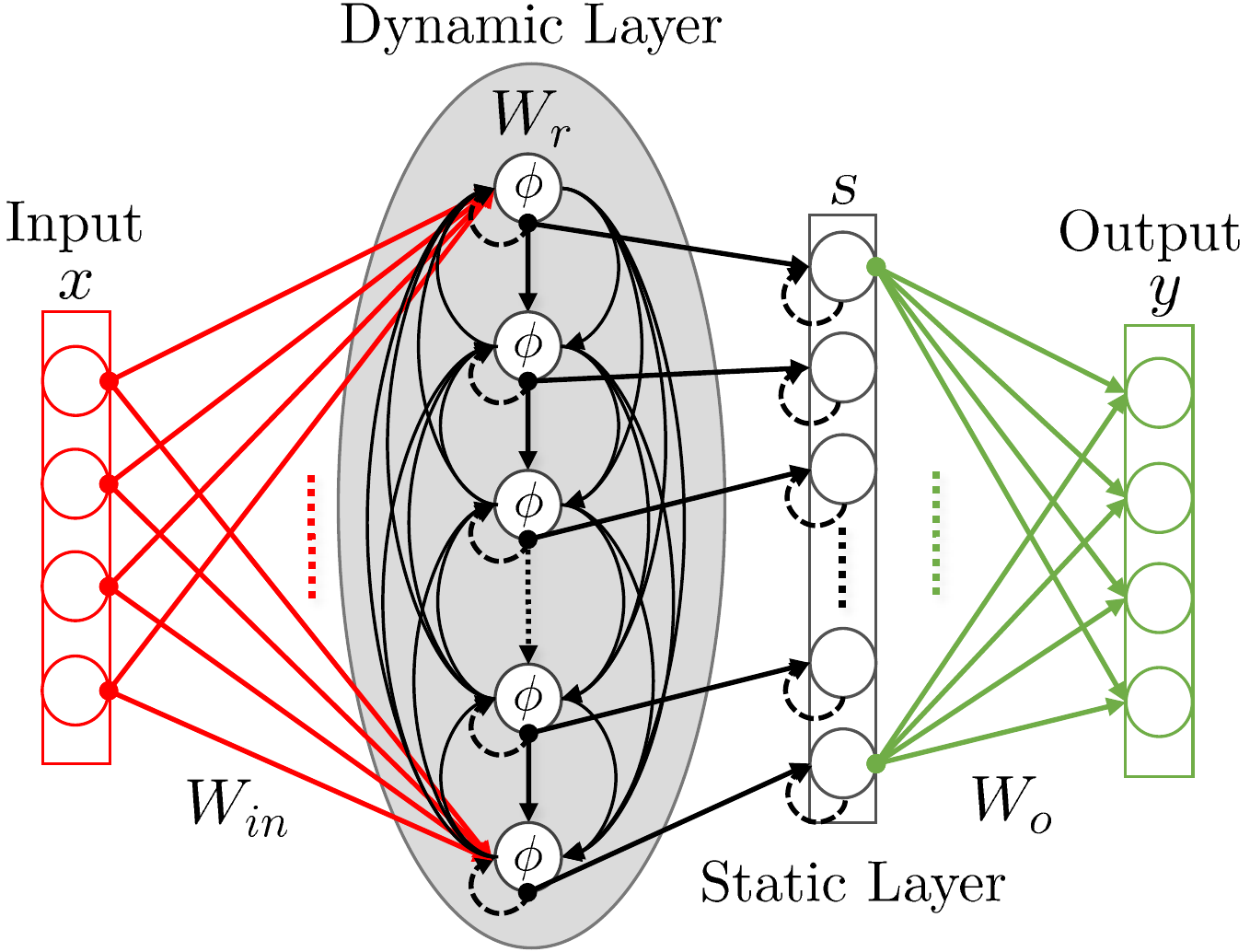}
\caption{Schematic of a leaky-ESN. A dynamical core called a reservoir is driven by input signal $x$. The states of the reservoir $s$ are combined non-linearly to produce the output $y$. The reservoir consists of $N$ interconnected nodes followed by a static (leaking) layer. The circular dashed lines in dynamic and static layers represent the past state values, while the solid lines represent the current time step value. The reservoir and input weights are fixed after initialization, while the output weights are learned using a regression technique.}
\label{ESNFig}
\end{figure}

The ESN is a promising type of RNNs due to its relaxed training complexity and its ability to preserve the temporal features from different signals over time~\cite{verzelli2019echo,sun2020review, ranzini2021experimental, sorokina2020dispersion,ren2020performance}. The ESNs are in the reservoir computing (RC) category because in the ESNs only the output weights are trainable. In Fig.~\ref{ESNFig}, the grey-colored area is the reservoir ``main'' structure containing the randomly connected ``neurons'' that capture the time features of the signal, while the output weights are trained to define which states are more relevant to describe the desired output. 
In this paper, we use the concept of leaky-ESN~\cite{wu2018statistical} containing no output feedback connections. Our motivation to choose the leaky-ESN architecture is that there was an experimental observation that the leaky-ESN configuration outperforms the traditional ESN in feature extraction for noisy time series~\cite{sun2020review}. The latter is, evidently, an important property in optical transmission-related tasks. The leaky-ESN is formalized for a certain time-step $t$, as follows: 
\begin{equation}\label{eq.ESN1}
 a_t = \phi \left( W_{r} \times s_{t-1} + W_{\text{in}} \times x_t \right),
\end{equation}
\begin{equation}\label{eq.ESN2}
  s_t = (1- \mu) s_{t-1} + \mu a_t,
\end{equation}
\begin{equation}\label{eq.ESN3}
y_t = W_{o} \times s_t,
\end{equation}
where $s_t \in \mathbb{R}^{N_r}$ is the system state at time-step $t$, $N_r$ is the number of hidden neurons units in the dynamic layers, which represents the dimensionality in the reservoir; $x_t \in \mathbb{R}^{n_i}$ and $y_t \in \mathbb{R}^{n_o}$ are the input and the output vector of the ESN, respectively; $W_r \in \mathbb{R}^{N_r \times N_r}$ is a reservoir weight matrix that defines which neuron units are connected (including the self-connections); this matrix is also characterized by a sparsity parameter $s_p$ defining the ratio of connected neurons to the total possible connections number. Finally, $W_{in} \in \mathbb{R}^{N_r \times n_i}$ is the input weight matrix, $\mu$ is the leaking rate parameter, and $W_o \in \mathbb{R}^{n_o \times N_r}$ is the output weight matrix which is the only one that is trainable using a regression technique. This training phase in ESN does not affect the dynamics of the system, which makes it possible to operate with the same reservoir for different tasks~\cite{verzelli2019echo}. A schematic representation of a leaky-ESN, including the sequential input, dynamic, static, and output layers, as depicted in Fig.~\ref{ESNFig}. 
 
The signal passing through the dynamic layer in Fig.~\ref{ESNFig} is represented by (\ref{eq.ESN1}), and this layer is the core of the reservoir structure. Then it is followed by a static layer, represented by (\ref{eq.ESN2}), which incorporates the leaky-ESN behavior through accumulating (integrating) its inputs, but it is also losing exponentially (leaking) accumulated excitation over time. Finally, the output layer defines which units are relevant to the description of the current task (for the equalization, in our case), and it is described by (\ref{eq.ESN3}). 
 
Concerning the previous ESN applications for optical channel equalization~\cite{sorokina2020dispersion}, the ESN was implemented in the optical domain for the distortions' mitigation: a 2~dB gain in $Q^2$-factor was achieved for 64-QAM 30~GBaud signals transmitted through 100~km fiber at 10~dBm input power. In addition, the same as it is in our paper, the reservoir can be applied in the digital domain. In~\cite{ranzini2021experimental}, the leaky-ESN was successfully applied after the analog-to-digital converter to enable 80 km transmission to reach below KP4-FEC limit~\cite{el2019400} for a 32 GBd on-off keying signal. 

\subsection{Convolutional neural networks} \label{subsec:CNN}

Due to their feature extraction propensity~\cite{aggarwal2018neural}, the CNNs have become one of the most commonly used NN structures in such areas as 2D image classification and 3D video applications~\cite{ciresan2011flexible,karpathy2014large}. Convolution layers have also been found efficient in the analysis of temporal 1D sequences with several applications to time series sensors, audio signals, and natural language processing~\cite{gu2018recent,lecun1995convolutional}. For longer sequences, the CNN layer can be used as a \textit{pre-processing step} due to its ability to reform the original sequence and extract its high-level features used for further processing cycles~\cite{bajaj2020single}.

Here, we investigate, for the first time, two new models for the equalization of signal distortions in metro systems, combining a 1D convolutional layer performing the effective signal pre-processing with two previously proposed NN-based equalizers: the MLP described in Sec.~\ref{subsec:MLP}, and  the biLSTM, Sec.~\ref{subsec:biLSTM}. These new structures, CNN+biLSTM and CNN+MLP, are addressed in our study because it was shown that the convolution layers are efficient in image denoising~\cite{zhang2017beyond} and array signal processing~\cite{zhao2018convolutional}, where the CNNs can reduce the background and quantization noise effects on coded signals. Therefore, we can naturally surmise that in our model the first convolutional layer can enhance the received signal by removing a part of the embedded noise before it enters the next neural layer. Also, generally, by adding the CNN layer, we end up with a NN model with less trainable parameters without losing performance, which can be yet another advantage. To that end, in the current study, we analyze how the combined NN architectures work for the optical channel equalization task. A simplified CNN+MLP combination was already successfully used in~\cite{bajaj2020single} at the transceiver for the high-baud-rate 80 km system.

The convolutional layer is primarily characterized by three key parameters: the number of filters, the size of its kernel, and the layer activation function. The extracting functionality is achieved by applying $n_f$ filters, sliding over the raw input sequence, and generating the number of output maps equal to $n_f$, with a fixed kernel size $n_k$. The convolutional layer is constructed as a squash function, which means that the input is mapped to a lower-dimensional representation, in which only the main (or desirable) characteristics are retained. Since the CNNs were mainly developed in the context of image recognition and spacial feature extraction, other parameters such as padding, dilation, and stride, are also used in the design of the convolutional layers. Considering that the input shape is $(B,M,4)$, the output shape after the CNN layer with all those parameters is defined as $(B,L_{out},n_f)$, where the parameter $L_{out}$ is the function of the CNN hyper-parameters and defined as:
\begin{equation}\label{eq.Lout}
  L_{out} \! = \! \left[ \frac{M +2 \! \cdot \! padding \! - \! dilation \! \cdot \! (n_k-1) \! - \! 1 }{stride} \! + \! 1\right].
\end{equation}
However, in this paper, we will not focus on the investigation of those additional parameters. Consequently, we fix the default convolutional layer configuration with the padding equal to $0$ (which corresponds to ``valid'' in Keras), the dilation equal to $1$, and the stride equal to $1$. Then, the input-output mapping of the convolution layer for this configuration can be described as follows:
\begin{equation}\label{eq.CNN}
  y^{f}_{i} = \phi \left(\sum_{n=1}^{n_i}\sum_{j=1}^{n_k}x^{in}_{i+j-1,n} \odot k^{f}_{j,n} + b^{f}_{j,n} \right),
\end{equation}
where $y^{f}_{i}$ is the output feature map of the $i$-th input element produced by filter $f$ in the CNN layer, \(x^{in}\) is the input raw data vector, $k^{f}_{j}$ is the \textit{j}-th trained convolution kernel of the filter $f$, and \(b^{f}_{j}\) is the bias of the filter $f$. Further, $n$ is the feature index of the kernel and input data, ranging from 1 to $n_i$, corresponding to the number of features in the data; $\phi$, as before, denotes the nonlinear activation function used in the convolutional layer. Note that (\ref{eq.CNN}) is true for all $i \in [1,..., L_{out}]$. Moreover, since the pooling layer captures only the most important features in the data and ignores the less important ones~\cite{hassan2018convolutional}, the pooling discretization process is not used in our equalizers to avoid the downsampling of feature sequences.

The output collection of feature maps, \(y^{f}\), emerging from the convolutional layer, is then fed into one of the structures described above: either into two dense layers (MLP, where the number of layers is, again, dictated by the BO), forming the CNN+MLP structure or into the one biLSTM layer, resulting in CNN+biLSTM. We recall that we use the convolutional layer before the following layers to extract the middle-level locally invariant features from the input series.

Here we mention that even the CNNs alone are extremely powerful deep learning instruments that have a complicated multiparametric structure that combines filters, kernel size, padding, stride, dilation, and pooling. However, having performed an exhaustive experimental exploration, we observed that deep CNNs have not reached the substantial performance level, like the one achieved by CNN+ MLP or CNN+biLSTM in our test case. Therefore, in this work, we utilize the convolution layers as pre-processing feature-extracting step and do not include deep CNN architectures in our current study.

\begin{figure}[ht!]
\centering\includegraphics[width=8.9cm]{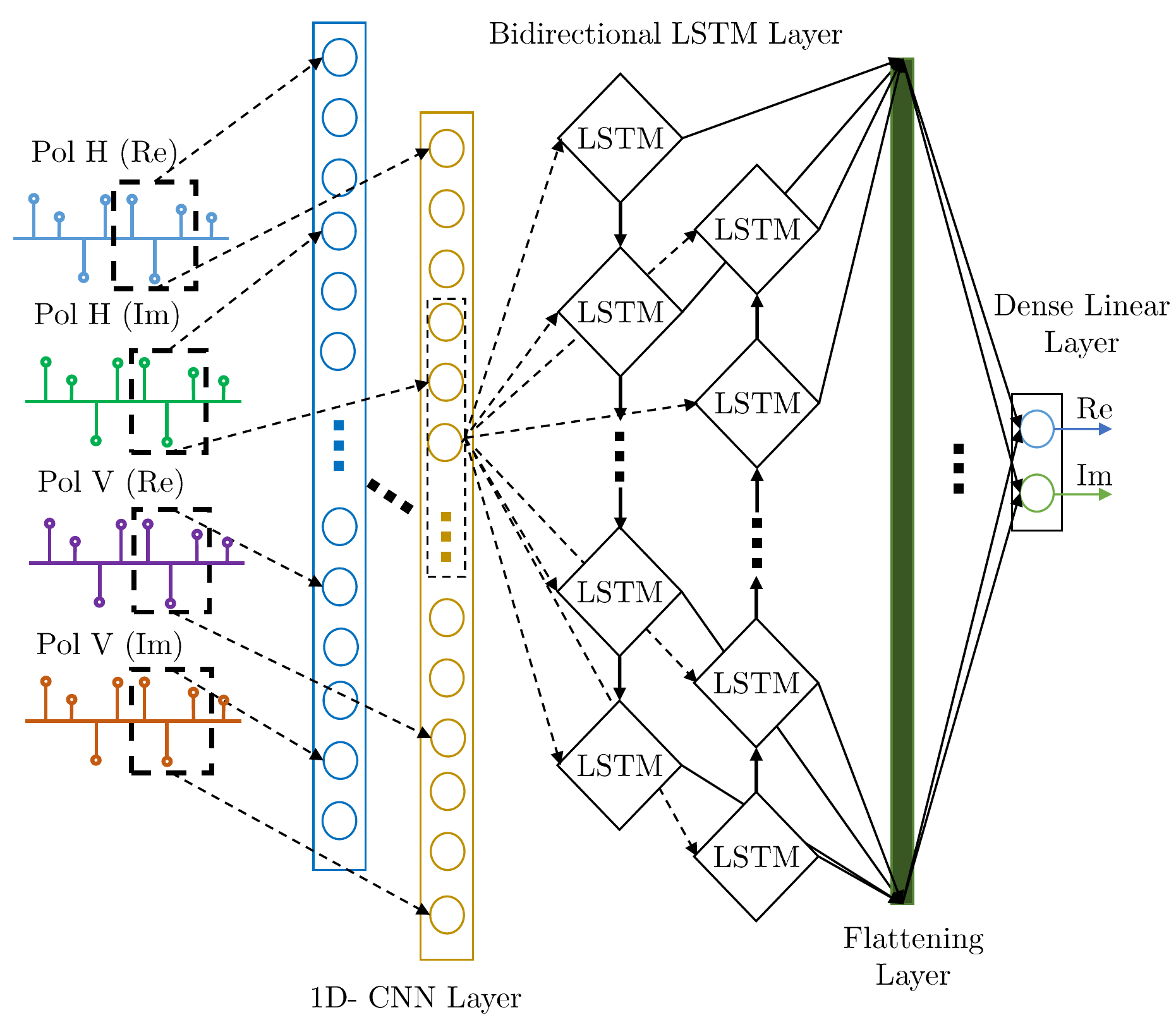}
\caption{Schematic of the convolutional-recurrent NN composed of an input layer corresponding to a time series with 4 features, followed by a 1D- convolutional layer (represented with two rectangles containing neurons) and a biLSTM layer, shown with the two lines (bi-directional) of lozenges, and ending with the flattering layer (a thick vertical deep-green line) and the output layer consisting two linear neurons to represent the real and imaginary part of the recovered symbol. }
\label{cnnmodel}
\end{figure}

\section{Computational complexity of the NN-based equalizers \label{Sec:complexity}}

In this section, the computational complexity in terms of real multiplications per recovered output symbol is examined for all introduced NN architectures. We notice that the number of additions is typically neglected for such estimation in ordinary DSP techniques~\cite{spinnler2010equalizer}. The major reason for this is that the typical algorithms for multiplying two integers with $n$ digits have a computational complexity of $\mathcal{O}(n^2)$, whereas adding the same two numbers has a computational complexity of $\Theta(n)$ \cite{jahani2009zot}. As a result, due to dealing with float values with 16 decimal digits, multiplication is by far the most time-consuming part of the implementation procedure. 

Here we point out that the training complexity will not be considered since we evaluate the real-time computation complexity (evaluation phase), which is the most critical part, while the training of a NN equalizer is carried out offline (calibration phase). Also, the computational complexity of nonlinear activation functions is not considered in our framework, due to the fact that typically their operation is based on an approximation approach,  rather than on direct multiplicative calculation. In the classical lookup tables-based  (LUTs)  approximation method, direct mapping can be digitally implemented with much fewer computations required to apply such activation functions~\cite{716806, 9170828}.

Early works presented the results regarding the complexity of the MLP~\cite{sidelnikov2018equalization}, RNN~\cite{zhou2019low}, and LSTM~\cite{lu2019memory} layers. However, to enhance the understanding of this subject and clarify it, in our work we directly relate those complexities to the parameters of the most widely used machine learning platforms (Keras, TensorFlow, and PyTorch) without losing generality, and specifically addressing the composite NN types described before. 
Let the mini-batch size be $B$, $n_s$ be the input time sequence size, with $n_s=M$, where $M$ is the memory size (see also Sec.~\ref{sec:zoo}), and $n_{i}$ be the number of features, which in our case is equal to $4$. Since we recover the real and imaginary parts of each symbol, the number of outputs per symbol, $n_o$, is equal to $2$. For ESN, biLSTM, and CNN layers, as they require inputs in the form of tensors of rank 3, the input of the NN equalizer can be parametrized as $[B, n_s, n_i]$, the three numbers defining the dimensions of the input tensor, as mentioned above. The parametrization for the MLP equalizer is simpler, with $[B,n_s \cdot n_i]$ defining the dimensions of the 2D tensor input. We use flattening layers when it was necessary to reduce the dimensionality of the data. 

In this case, considering three dense layers with $n_1$, $n_2$, and $n_3$ neurons, respectively, the complexity $C_{\text{MLP}}$ of the resulting NN is given by:
\begin{equation}
\label{Eq_c1}
  C_{\text{MLP}}=
  \underbrace{n_s n_i n_1}_{\text{$a_1$}}+
  \underbrace{n_1n_2+ n_2n_3}_{\text{$b_1$}}+
  \underbrace{n_3n_o}_{\text{$c_1$}},
\end{equation}
where $a_1$ is the contribution of the input layer, $b_1$ is the contribution of the hidden layer, and $c_1$ refers to the contribution of the output layer. The subindex ``1'' in $a$, $b$, and $c$ explicitly associates these parameters with the MLP architecture

The next part presents the computational complexity for an NN-based equalizer composed of a biLSTM layer.
Assuming that the biLSTM layer has $n_{h}$ hidden units, the complexity of such a NN is given by:
\begin{equation}
\label{Eq_c2}
\begin{split}
  C_{\text{biLSTM}}= 2
  \underbrace{n_{s}n_{h}(4n_i+4n_{h}+3+n_o)}_{\text{$a_2$}},
  \end{split}
\end{equation}
where $a_2$ is the contribution of the only layer, while the subindex ``2'' attributes the number $a$ to the biLSTM. This expression is easier to understand if we analyze the mathematical description of the LSTM cell, see (\ref{EQ.LSTM}) and Fig.~\ref{LSTMcell}. We have several contributions to the cell's complexity. In the first layer we have $4n_{i} n_{h}$ multiplications associated with the input vector $x_t$. Then, $4n_{h}^2$ multiplications are due to the operations with the previous cell output $h_{t-1}$. Afterward, $3n_{h}$ and $n_o n_{h}$ multiplications due to the internal multiplications identified with $\odot$ and involving the current cell output ($h_t$) going into the output layer, respectively, are added. Lastly, we multiply the number of operations by the number of time steps in the layer, $n_s$. Since the topology is bidirectional, the total contribution is also multiplied by $2$. 

Following Sec.~\ref{sec:zoo}, now we address the computational complexity associated with the ESN equalizer. Before presenting the respective expression, it is important to emphasize two aspects. First, the implementation of the ESN in the digital domain does not benefit from the fact that only the output layer weights are trainable, since, as mentioned previously, the training is not a key bottleneck as it is carried out during the offline calibration process. Second, the complexity of the ESN can potentially drop drastically if we implement it in the optical domain as an ESN dynamic layer, as it was noted in~\cite{sorokina2020dispersion}. However, in this paper, we analyze the ESN implementation in the digital domain, similarly to~\cite{ranzini2021experimental}.  

Considering the leaky-ESN definition given by (\ref{eq.ESN1})--(\ref{eq.ESN3}), the computational complexity of this equalizer can be expressed as:
\begin{equation}
\label{Eq_c5} \begin{split}
    C_{\text{ESN}}= n_s\left(
   \underbrace{n_iN_r + N_{r}^{2}s_p}_{\text{$a_3$}}+
   \underbrace{2N_r}_{\text{$b_3$}} + \underbrace{N_r n_o}_{\text{$c_3$}} \right).\end{split}
\end{equation}\label{complexityESN}
In the expression above, $a_3$ represents the contributions of  (\ref{eq.ESN1}), where the input layer adds $n_iN_r$ multiplications whereas the dynamic layers add $N_r^{2}s_p$. $b_3$ refers to the contributions of (\ref{eq.ESN2}) describing the static layer, and $c_3$ represents the multiplications in the output layer, (\ref{eq.ESN3}). This overall process is repeated for all $n_s$ time steps. Note that in the case of a potential optical implementation of the ESN, $a_3$ and $b_3$ would be equal to zero, and only the final weights would be learned in the digital domain.

Finally, let us address the complexity of the composite structures: CNN+MLP and CNN+biLSTM. The computational complexity of a 1-D convolutional layer is described as:

\begin{equation}
\label{Eq_conv_all}
    C_{\text{CNN}}= n_{i} n_{f} n_{k}\! \left[ \frac{n_s +2  \, padding \! - \! dilation (n_k-1) \! - \! 1 }{stride} \! + \! 1\right]
\end{equation}

However, we assumed (\ref{eq.CNN}) that the convolutional layer is defined by the number of filters $n_f$, the kernel size $n_k$, and that the number of time steps $n_s \geq n_k$ , according to (\ref{eq.Lout}) the output size for each filter of the CNN is $(n_s - n_k + 1)$. (\ref{Eq_c3}) and~(\ref{Eq_c4}) are the expressions for the complexity of a convolutional layer combined with two dense layers or one biLSTM layer, respectively: 
\begin{equation}
\label{Eq_c3}\begin{split}
    C_{\text{CNN+MLP}}= 
   \underbrace{\vphantom{3M_{\rm{Volt}}^3}n_in_fn_k(n_s-n_k+1)}_{\text{$a_4$}}+ \\
   \underbrace{\vphantom{3M_{\rm{Volt}}^3}(n_s-n_k+1)n_f}_{\text{$b_4$}} \underbrace{n_1 +n_1n_2+n_2n_o}_{\text{$c_4$}} \, , \end{split}
\end{equation}
\begin{equation}
\label{Eq_c4} \begin{split}
    C_{\text{CNN+biLSTM}}= 
   \underbrace{\vphantom{2n_h[4n_f+4n_h+3+n_o]}n_in_fn_k(n_s-n_k+1)}_{\text{$a_5$}}+\\
   \underbrace{\vphantom{2n_h[4n_f+4n_h+3+n_o]}(n_s-n_k+1)}_{\text{$b_5$}} \underbrace{2n_h[4n_f+4n_h+3+n_o]}_{\text{$c_5$}}.\end{split}
\end{equation}
In this scenario, the two-layer MLP has $n_1$ and $n_2$ neurons in each respective layer, and the biLSTM layer has $n_h$ hidden units. In the equations above, $a_4$ and $a_5$ are the contributions of the convolutional layer, $b_4$ is the correction factor for the transition between layers since the flattening layer was placed before the dense layers; $b_5$ is the number of time-steps for the following biLSTM layer; $c_4$ is the contribution of the two-layer MLP; and $c_5$ is the contribution of the biLSTM layer where, in this case, the number of filters, $n_f$, is equal to the number of features entering the LSTM cell. \par

Finally, we would like to express the computational complexity of the DBP-based receiver used in this paper for benchmark purposes. We considered a basic implementation of the DBP algorithm \cite{lin2014adaptive}, where each propagation step comprises a linear part for dispersion compensation followed by a nonlinear phase cancellation stage. The linear part is achieved with a zero-forcing equalizer by transforming the signal in the frequency domain and multiplying with the inverse dispersion transfer function of the propagation section. The complexity of the DBP in terms of RMpS is~\cite{spinnler2010equalizer, sidelnikov2018equalization}:

\begin{equation}\label{eq:eq.V}
\begin{split}
C_\text{DBP}\! = \! 4N_{span}N_{step} \! \left(\frac{n \, N_\text{FFT} [\log_2(N_\text{FFT})+1]}{(N_\text{FFT}-N_D +1)}+n \! \! \, \right)\!,
\end{split}
\end{equation}
where $N_{step}$ is the number of steps per span used, $N_\text{FFT}$ is the FFT size, $n$ is the oversampling ratio, and $N_D=\tau_D/T $, where $\tau_D$ corresponds to the dispersive channel impulse response and $T = 1/R_s$ is the symbol duration. We have considered that $N_\text{FFT} = 256$ and $\tau_D$ defined as:

\begin{equation}\label{eq:eq.tald}
\tau_D = \frac{1.1 R_{s} c |D| L_{span} }{f_{c}^{2} N_{steps}},
\end{equation}
where $f_{c}$ is the optical carrier reference frequency that in our case was $193.41$ THz, $c$ is the speed of light, $L_{span}$ is the span length and $D$ is the fiber dispersion parameter.

\section{Performance versus computational complexity trade-off analysis} \label{Sec:results}
In this section, we initially describe the numerical and experimental scenarios used in this paper to analyze and compare the functioning of the equalizers detailed in Sec. \ref{sec:zoo}. After that, the two types of analysis for our set of NN structures are carried out. First, we present the maximum performance improvement (in terms of Q-factor gain compared to the non-equalized case) that each equalizer can deliver and compare this gain to the respective computational complexity corresponding to each optimized equalizer. Then, we decrease the computational complexity of six NN topologies from Sec.~\ref{sec:zoo} and present the gain improvement provided by each NN-equalizer when all NNs have approximately the same computational complexity. This enables us to investigate the dependence of optical performance on the computational complexity and to identify which equalizer is better for a certain complexity level.

\subsection{Experimental and numerical setups}\label{subsec:exp}

\begin{figure}[ht!]
\centering\includegraphics[width=7.0cm]{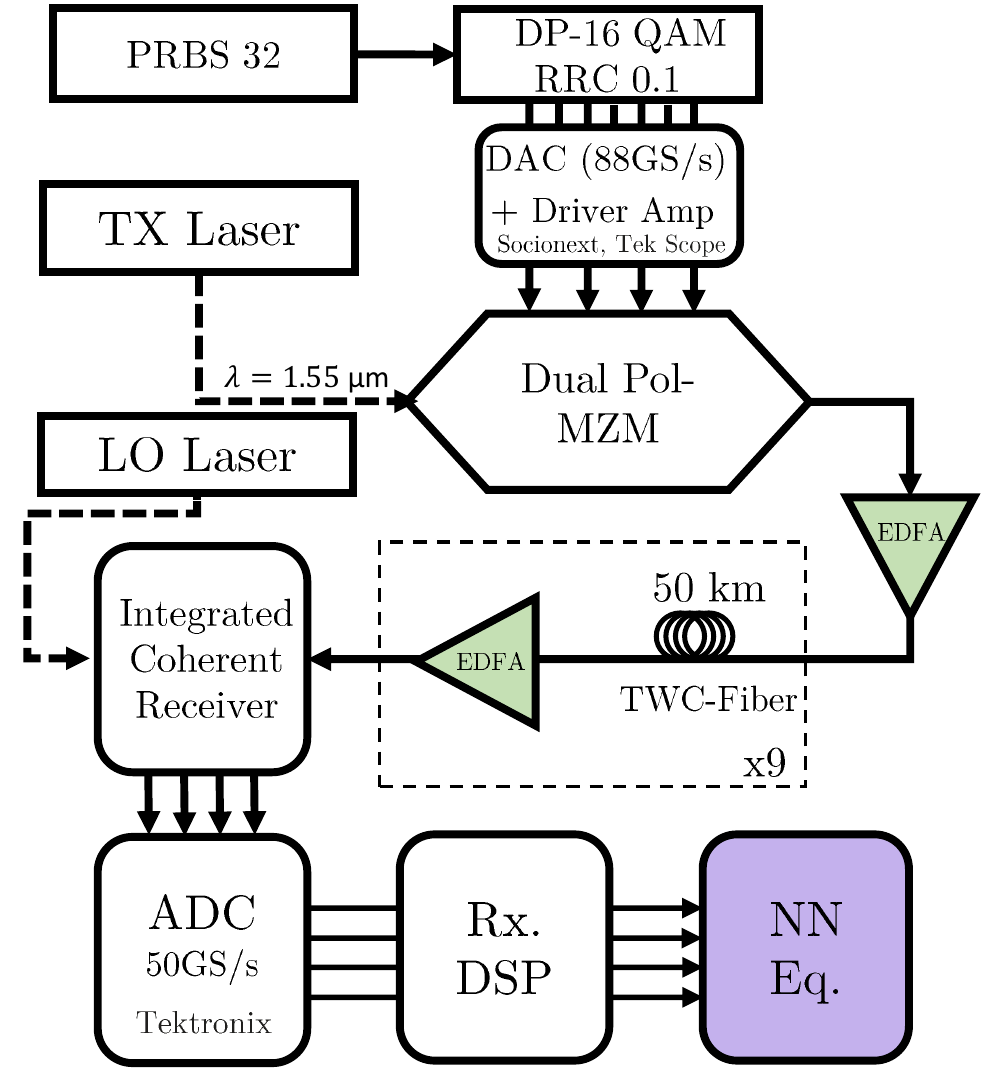}
\caption{Experimental setup used to analyze the performance of different NN equalizers; further details are reported in Sec.~\ref{subsec:exp}. The input of the NN (shown as the purple rectangle at the bottom right) is the soft output of the regular DSP just before the decision unit.}
\label{setup}
\end{figure}

The setup used in our experiment is depicted in Fig.~\ref{setup}. At the transmitter, a DP-16QAM 34.4 Gbaud symbol sequence was mapped out of data bits generated by a $2^{32} - 1$ PRBS. Then, a digital RRC filter with roll-off 0.1 was applied to limit the channel bandwidth to 37.5 GHz. The resulting filtered digital samples were resampled and uploaded to a digital-to-analog converter (DAC) operating at 88 Gsamples/s. The outputs of the DAC were amplified by a four-channel electrical amplifier which drove a dual-polarization in-phase/quadrature Mach–Zehnder modulator, modulating the continuous waveform carrier produced by an external cavity laser at $\lambda = 1.55 \mu m$. 
The resulting optical signal was transmitted over 9$\times$50 km spans of TWC optical fiber with EDFA amplification. The optical amplifier noise figure was in the 4.5 to 5 dB range. The parameters of the TWC fiber -- at $\lambda = 1.55 \mu m$ -- are: attenuation coefficient $ \alpha = 0.23$ dB/km, dispersion coefficient $D = 2.8$ ps/(nm·km), and effective nonlinear coefficient $\gamma$ = 2.5 ($W\cdot km)^{-1}$. 

At the RX side, the optical signal was converted into the electrical domain using an integrated coherent receiver. The resulting signal was sampled at 50 Gsamples/s by a digital sampling oscilloscope and processed by an offline DSP based on the algorithms described in~\cite{kuschnerov2010data}. Firstly, the bulk accumulated dispersion was compensated using a frequency domain equalizer, which was followed by the removal of the carrier frequency offset. A constant-amplitude zero-autocorrelation-based training sequence was then located in the received frame and the equalizer transfer function was estimated from it. After the equalization, the two polarizations were demultiplexed and the signal was corrected for clock frequency and phase. Carrier phase estimation was then achieved with the help of pilot symbols. Thereafter, the resulting soft symbols were used as input for the NN equalizers. Finally, the pre-FEC BER was evaluated from the signal at the NN output.

\begin{figure*}[ht!] 
  \begin{subfigure}[b]{0.5\linewidth}
    \centering
\begin{tikzpicture}[scale = 0.74]
    \begin{axis}[
        my axis style,
        ymin=0,
        ymax=6,
        axis y line*=left,
        ylabel = {\textcolor{red}{Q-factor Gain [dB]}},
        % to avoid overlapping, move the "first" bars to the left ...
        bar shift={-\pgfplotbarwidth/2},
    ]
        \addplot [style={red,fill=red,mark=none}] coordinates {
            (MLP,2.98)
            (biLSTM, 4.33)
            (ESN, 0.42)
            (CNN+MLP, 3.91)
            (CNN+biLSTM, 4.38)
            (DBP, 4.12)
        };
    \end{axis}
    \begin{axis}[
        my axis style,
        ymin=0,
        ymax=9,
        axis x line=none,
        axis y line*=right,
      ylabel = {\textcolor{blue}{ $log_{10}$(RMpS)}},
        % ... and the second bars to the right
        bar shift={\pgfplotbarwidth/2},
    ]
        \addplot [style={blue,fill=blue,mark=none}] coordinates {
            (MLP, 5.09)
            (biLSTM, 7.23)
            (ESN, 4.93)
            (CNN+MLP, 6.89)
            (CNN+biLSTM, 7.52)
             (DBP, 3.334412446)

        };
    \end{axis}
\end{tikzpicture}
    \caption{Simulation results}
    \label{numerical_result1} 
  \end{subfigure}%%
  \begin{subfigure}[b]{0.5\linewidth}
    \centering
\begin{tikzpicture}[scale = 0.74]
    \begin{axis}[
        my axis style,
        ymin=0,
        ymax=4,
        axis y line*=left,
        ylabel = {\textcolor{red}{Q-factor Gain [dB]}},
        % to avoid overlapping, move the "first" bars to the left ...
        bar shift={-\pgfplotbarwidth/2},
    ]
        \addplot [style={red,fill=red,mark=none}] coordinates {
            (MLP,1.95)
            (biLSTM, 2.76)
            (ESN, 0.58)
            (CNN+MLP, 2.2)
            (CNN+biLSTM, 2.91)
            (DBP, 1.32)
        };
    \end{axis}
    \begin{axis}[
        my axis style,
        ymin=0,
        ymax=9,
        axis x line=none,
        axis y line*=right,
      ylabel = {\textcolor{blue}{ $log_{10}$(RMpS)}},
        % ... and the second bars to the right
        bar shift={\pgfplotbarwidth/2},
    ]
        \addplot [style={blue,fill=blue,mark=none}] coordinates {
            (MLP, 5.09)
            (biLSTM, 7.23)
            (ESN, 4.93)
            (CNN+MLP, 6.89)
            (CNN+biLSTM, 7.52)
            (DBP, 3.334412446)
        };
    \end{axis}
\end{tikzpicture}
    \caption{Experimental results}
    \label{experiment_result1} 
  \end{subfigure} 
  
  \caption{Comparison of the computational complexity versus performance for the different NN-based equalizer considered within this paper with their optimized architectures and the DBP with 3 StPS. The number over each bar gives the 10 logarithm of the number of multiplications per recovery symbol.}
  \label{Result_01} 
\end{figure*}
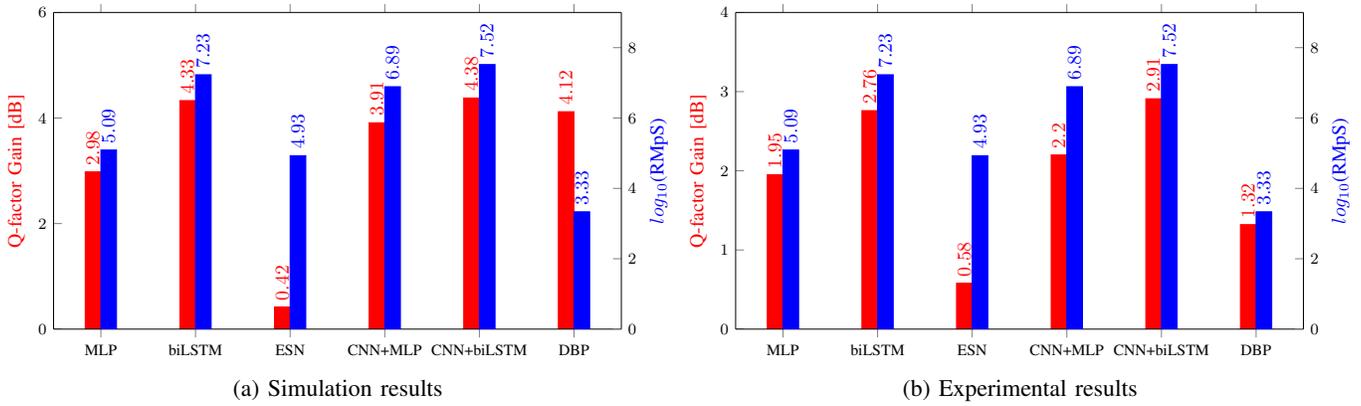

\begin{table}[h!]
\centering
\caption{Summary of the complexity attributing to each NN equalizer topology: the topology type is identified in the leftmost column. The complexity corresponding to each topology and the NN type is expressed in terms of real multiplications per symbol recovered (RMpS), highlighted in red. In this table, we also depict the hyper-parameters distributions found by the BO: the cell marked as ``Best topology'' and other 6 topologies (Topologies from 1 to 6, referring to the increasing complexity threshold number) for the study of complexity versus performance. In addition, for all topologies, the values of $n_s$, $n_i$ and $n_o$ were: $41$, $4$ and $2$, respectively, and these are not reported in the table.} 
\begin{adjustbox}{width=\textwidth/2, center}
\begin{tabular}{|c|c|c|c|c|c|c|c|c|c|}
\hline
 & \multicolumn{4}{c|}{\textbf{CNN+biLSTM}} & \multicolumn{2}{c|}{\textbf{biLSTM}} & \multicolumn{3}{c|}{\textbf{ESN}} \\ \cline{2-10} 
 & $n_f$ & $n_k$ & $n_h$ & {\color[HTML]{FE0000} RMpS} & $n_h$ & {\color[HTML]{FE0000} RMpS} & $N_r$ & $s_p$ & {\color[HTML]{FE0000} RMpS} \\ \cline{2-10} 
 & $244$ & $10$ & $226$ & {\color[HTML]{FE0000} $2.7$E+07} & $226$ & {\color[HTML]{FE0000} $1.7$E+07
} & $88$ & $0.18$ & {\color[HTML]{FE0000} $8.6$E+04} \\ \cline{2-10} 
 & \multicolumn{5}{c|}{\textbf{CNN+MLP}} & \multicolumn{4}{c|}{\textbf{MLP}} \\ \cline{2-10} 
 & $n_f$ & $n_k$ & $n_1$ & $n_2$ & {\color[HTML]{FE0000} RMpS} & $n_1$ & $n_2$ & $n_3$ & {\color[HTML]{FE0000} RMpS} \\ \cline{2-10} 
\multirow{-6}{*}{\textbf{ \rotatebox[origin=c]{90}{Best Topology}
}} & $470$ & $10$ & $456$ & $467$ & {\color[HTML]{FE0000} $7.7$E+06} & $149$ & $132$ & $596$ & {\color[HTML]{FE0000} $1.2$E+05}

\\ \hline \hline

 & \multicolumn{4}{c|}{\textbf{CNN+biLSTM}} & \multicolumn{2}{c|}{\textbf{biLSTM}} & \multicolumn{3}{c|}{\textbf{ESN}} \\ \cline{2-10} 
 & $n_f$ & $n_k$ & $n_h$ & {\color[HTML]{FE0000} RMpS} & $n_h$ & {\color[HTML]{FE0000} RMpS} & $N_r$ & $s_p$ & {\color[HTML]{FE0000} RMpS} \\ \cline{2-10} 
 & $1$ & $10$ & $1$ & {\color[HTML]{FE0000} $2.1$E+03} & $1$ & {\color[HTML]{FE0000} $2.0$E+03} & $6$ & $0.18$ & {\color[HTML]{FE0000} $2.2$E+03} \\ \cline{2-10} 
 & \multicolumn{5}{c|}{\textbf{CNN+MLP}} & \multicolumn{4}{c|}{\textbf{MLP}} \\ \cline{2-10} 
 & $n_f$ & $n_k$ & $n_1$ & $n_2$ & {\color[HTML]{FE0000} RMpS} & $n_1$ & $n_2$ & $n_3$ & {\color[HTML]{FE0000} RMpS} \\ \cline{2-10} 
\multirow{-6}{*}{\textbf{ \rotatebox[origin=c]{90}{Topology 1}}} & $2$ & $5$ & $10$ & $10$ & {\color[HTML]{FE0000} $2.3$E+03} & $10$ & $10$ & $25$ & {\color[HTML]{FE0000} $2.0$E+03} 

\\\hline \hline

 & \multicolumn{4}{c|}{\textbf{CNN+biLSTM}} & \multicolumn{2}{c|}{\textbf{biLSTM}} & \multicolumn{3}{c|}{\textbf{ESN}} \\ \cline{2-10} 
 & $n_f$ & $n_k$ & $n_h$ & {\color[HTML]{FE0000} RMpS} & $n_h$ & {\color[HTML]{FE0000} RMpS} & $N_r$ & $s_p$ & {\color[HTML]{FE0000} RMpS} \\ \cline{2-10} 
 & $5$ & $10$ & $3$ & {\color[HTML]{FE0000} $1.3$E+04} & $4$ & {\color[HTML]{FE0000} $1.2$E+04} & $22$ & $0.18$ & {\color[HTML]{FE0000} $1.1$E+04} \\ \cline{2-10} 
 & \multicolumn{5}{c|}{\textbf{CNN+MLP}} & \multicolumn{4}{c|}{\textbf{MLP}} \\ \cline{2-10} 
 & $n_f$ & $n_k$ & $n_1$ & $n_2$ & {\color[HTML]{FE0000} RMpS} & $n_1$ & $n_2$ & $n_3$ & {\color[HTML]{FE0000} RMpS} \\ \cline{2-10} 
\multirow{-6}{*}{\textbf{ \rotatebox[origin=c]{90}{Topology 2}}} & $9$ & $5$ & $12$ & $30$ & {\color[HTML]{FE0000} $1.1$E+04} & $40$ & $40$ & $80$ & {\color[HTML]{FE0000} $1.1$E+04} 

\\\hline \hline

 & \multicolumn{4}{c|}{\textbf{CNN+biLSTM}} & \multicolumn{2}{c|}{\textbf{biLSTM}} & \multicolumn{3}{c|}{\textbf{ESN}} \\ \cline{2-10} 
 & $n_f$ & $n_k$ & $n_h$ & {\color[HTML]{FE0000} RMpS} & $n_h$ & {\color[HTML]{FE0000} RMpS} & $N_r$ & $s_p$ & {\color[HTML]{FE0000} RMpS} \\ \cline{2-10} 
 & $20$ & $10$ & $10$ & {\color[HTML]{FE0000} $1.1$E+05} & $16$ & {\color[HTML]{FE0000} $1.1$E+05} & $100$ & $0.18$ & {\color[HTML]{FE0000} $1.1$E+05} \\ \cline{2-10} 
 & \multicolumn{5}{c|}{\textbf{CNN+MLP}} & \multicolumn{4}{c|}{\textbf{MLP}} \\ \cline{2-10} 
 & $n_f$ & $n_k$ & $n_1$ & $n_2$ & {\color[HTML]{FE0000} RMpS} & $n_1$ & $n_2$ & $n_3$ & {\color[HTML]{FE0000} RMpS} \\ \cline{2-10} 
\multirow{-6}{*}{\textbf{ \rotatebox[origin=c]{90}{Topology 3}}} & $50$ & $9$ & $30$ & $100$ & {\color[HTML]{FE0000} $1.1$E+05} & $170$ & $170$ & $300$ & {\color[HTML]{FE0000} $1.1$E+05} 

\\\hline \hline

 & \multicolumn{4}{c|}{\textbf{CNN+biLSTM}} & \multicolumn{2}{c|}{\textbf{biLSTM}} & \multicolumn{3}{c|}{\textbf{ESN}} \\ \cline{2-10} 
 & $n_f$ & $n_k$ & $n_h$ & {\color[HTML]{FE0000} RMpS} & $n_h$ & {\color[HTML]{FE0000} RMpS} & $N_r$ & $s_p$ & {\color[HTML]{FE0000} RMpS} \\ \cline{2-10} 
 & $50$ & $10$ & $41$ & {\color[HTML]{FE0000} $1.0$E+06} & $53$ & {\color[HTML]{FE0000} $1.0$E+06} & $350$ & $0.18$ & {\color[HTML]{FE0000} $1.0$E+06} \\ \cline{2-10} 
 & \multicolumn{5}{c|}{\textbf{CNN+MLP}} & \multicolumn{4}{c|}{\textbf{MLP}} \\ \cline{2-10} 
 & $n_f$ & $n_k$ & $n_1$ & $n_2$ & {\color[HTML]{FE0000} RMpS} & $n_1$ & $n_2$ & $n_3$ & {\color[HTML]{FE0000} RMpS} \\ \cline{2-10} 
\multirow{-6}{*}{\textbf{ \rotatebox[origin=c]{90}{Topology 4}}} & $300$ & $10$ & $70$ & $200$ & {\color[HTML]{FE0000} $1.1$E+06} & $600$ & $600$ & $900$ & {\color[HTML]{FE0000} $1.0$E+06}

\\\hline \hline

 & \multicolumn{4}{c|}{\textbf{CNN+biLSTM}} & \multicolumn{2}{c|}{\textbf{biLSTM}} & \multicolumn{3}{c|}{\textbf{ESN}} \\ \cline{2-10} 
 & $n_f$ & $n_k$ & $n_h$ & {\color[HTML]{FE0000} RMpS} & $n_h$ & {\color[HTML]{FE0000} RMpS} & $N_r$ & $s_p$ & {\color[HTML]{FE0000} RMpS} \\ \cline{2-10} 
 & $244$ & $10$ & $108$ & {\color[HTML]{FE0000} $1.0$E+07} & $172$ & {\color[HTML]{FE0000} $1.0$E+07} & $1150$ & $0.18$ & {\color[HTML]{FE0000} $1.0$E+07} \\ \cline{2-10} 
 & \multicolumn{5}{c|}{\textbf{CNN+MLP}} & \multicolumn{4}{c|}{\textbf{MLP}} \\ \cline{2-10} 
 & $n_f$ & $n_k$ & $n_1$ & $n_2$ & {\color[HTML]{FE0000} RMpS} & $n_1$ & $n_2$ & $n_3$ & {\color[HTML]{FE0000} RMpS} \\ \cline{2-10} 
\multirow{-6}{*}{\textbf{ \rotatebox[origin=c]{90}{Topology 5}}} & $600$ & $12$ & $500$ & $500$ & {\color[HTML]{FE0000} $1.0$E+07} & $2100$ & $2100$ & $2500$ & {\color[HTML]{FE0000} $1.0$E+07} \\ \hline \hline

 & \multicolumn{4}{c|}{\textbf{CNN+biLSTM}} & \multicolumn{2}{c|}{\textbf{biLSTM}} & \multicolumn{3}{c|}{\textbf{ESN}} \\ \cline{2-10} 
 & $n_f$ & $n_k$ & $n_h$ & {\color[HTML]{FE0000} RMpS} & $n_h$ & {\color[HTML]{FE0000} RMpS} & $N_r$ & $s_p$ & {\color[HTML]{FE0000} RMpS} \\ \cline{2-10} 
 & $400$ & $10$ & $455$ & {\color[HTML]{FE0000} $1.0$E+08} & $550$ & {\color[HTML]{FE0000} $1.0$E+08} & $3660$ & $0.18$ & {\color[HTML]{FE0000} $1.0$E+08} \\ \cline{2-10} 
 & \multicolumn{5}{c|}{\textbf{CNN+MLP}} & \multicolumn{4}{c|}{\textbf{MLP}} \\ \cline{2-10} 
 & $n_f$ & $n_k$ & $n_1$ & $n_2$ & {\color[HTML]{FE0000} RMpS} & $n_1$ & $n_2$ & $n_3$ & {\color[HTML]{FE0000} RMpS} \\ \cline{2-10} 
\multirow{-6}{*}{\textbf{ \rotatebox[origin=c]{90}{Topology 6}}} & $1000$ & $10$ & $2900$ & $2200$ & {\color[HTML]{FE0000} $1.0$E+08} & $7050$ & $7050$ & $7000$ & {\color[HTML]{FE0000} $1.0$E+08} \\ \hline

\end{tabular}
\end{adjustbox}
\label{table_hyperparameters}
\end{table}

With regard to simulation, we mimic the experimental transmission setup\footnote{We consider a DP-16QAM, single-channel signal at 34.4 Gbaud pre-shaped by an RRC filter with 0.1 roll-off transmissions with an upsampling rate of 8 samples per symbol (275.2 GSamples/s) over a system consisting of $9\!\times\!50$~km TWC-fiber spans}. The optical signal's propagation along optical fiber was simulated by solving the Manakov equations via the split-step Fourier method (with a resolution of $1$km per step). Every span was followed by an optical amplifier with noise figure $\text{NF}=4.5$~dB, which fully compensates fiber losses and adds amplified spontaneous emission noise. At the receiver, after full electronic chromatic dispersion compensation (CDC) by the frequency-domain equalizer and downsampling to the symbol rate, the received symbols were normalized to the transmitted ones. Finally, we added Gaussian noise to the signal representing an additional transceiver distortion that we may have in the experiment, such that the Q-factor level of the simulated data matched the experimental one. The system performance is evaluated in terms of the Q-factor, defined as: $Q = 20 \: \mathrm{log_{10}} \left[\sqrt{2} \: \mathrm{erfc^{-1}}(2\,\rm BER)\right]$. 

\subsection{Optimized NN-based architectures}
In this section, we show the maximum achievable Q-factor for all equalizers without constraining the computational complexity. The Bayesian optimization (BO) tool, introduced in~\cite{freire2020complex} for optical NN-based equalizers, was implemented to identify the optimum values of hyper-parameters for each NN topology, which provides the best Q-factor in the experimental test dataset. As it was recently shown, the BO render superior performance compared to other types of search algorithms for machine learning hyperparameter tuning \cite{turner2021bayesian}. The same topologies (without further optimization) were tested for the numerical analysis as well. The search space used in the BO procedure was defined via the allowed hyper-parameters intervals:

$N=$[$1$ to $50$], $n_f=$[$1$ to $1000$],
$n_k=$[$1$ to $20$], $n_h=$[$1$ to $1000$], $n_1=$[$1$ to $1000$], $n_2=$[$1$ to $1000$], $n_3=$[$1$ to $1000$], $N_r=$[$1$ to $1000$], $s_p=$[$0$ to $1$], $\mu=$[$0$ to $1$], and spectral radius=[$0$ to $1$] .
 
In Table.~\ref{table_hyperparameters}, the line marked with the ``\textbf{Best Topology}'' label, summarizes the hyper-parameters obtained by BO. 
These values are used to count the real multiplications per symbol recovery (complexity), and to assess the equalizers' performance expressed via the Q-factor gain, Fig.~\ref{Result_01}. Note that for all equalizers, the same optimal number of taps found by the BO was $N=20$, which means that the memory in our equalizers is $M=41$ and the mini-batch size, $B$, is equal to 4331. Moreover, for the ESN, the BO found the best value $\mu = 0.57$, and the optimal spectral radius equal to 0.667. The activation functions found for every hidden NN layer are summarized as following: 1D-CNN layer – $‘linear’$ activation function followed by $LeakyReLU$ (Leaky version of a Rectified Linear Unit) with negative slope coefficient alpha = 0.2; biLSTM layer – hyperbolic tangent ($‘tanh’$) activation function; ESN layer – $‘tanh’$ activation function; MLP layer – $‘tanh’$ activation function.

The results obtained by using the numerical synthetic data are presented in Fig.~\ref{numerical_result1}. First, the CNN+biLSTM turned out to be best-performing in terms of the Q-factor gain: it achieved a 4.38 dB Q-factor improvement when compared to the conventional DSP algorithms~\cite{kuschnerov2010data}, 0.05~dB when compared to the biLSTM equalizer level, 0.47~dB when compared to the CNN+MLP equalizer level, 1.4~dB when compared to the MLP equalizer level, and 3.96 dB when compared to the ESN equalizer level. Second, when adding the convolutional layers to MLP and biLSTM, we observed the improvement in terms of the number of epochs needed to reach the highest performance: the single-layer biLSTM required 119 epochs, while the CNN+biLSTM reduced this number to 89 epochs; the MLP itself needed 214 epochs to reach the best performance level, and the CNN+MLP required just 100 epochs. Thus, we conclude that the addition of a convolutional layer indeed renders the enhancement in the NN structure's performance and assists in the training stage.

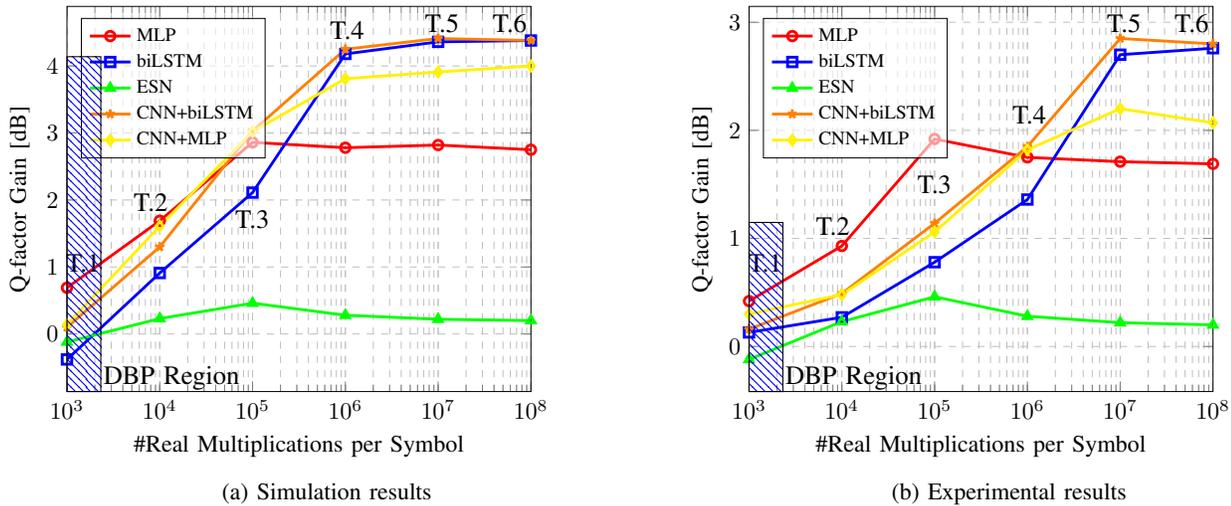
\begin{figure*}[ht!] 
  \begin{subfigure}[b]{\linewidth/2}
    \centering
 \begin{tikzpicture}[scale=0.9]
    \begin{axis} [ylabel={Q-factor Gain [dB]}, 
        xlabel={\#Real Multiplications per Symbol},
        grid=both,
        xmode=log, 
        xmin=1e3,
        xmax=1e8,
    	    ymajorgrids = true,
    grid style = dashed,
    % 	xtick={1, ..., 20},
    % 	ymin=0.00001, ymax=0.1,
    %   ymode=log,
    %   log ticks with fixed point,
        legend style={legend pos=north west, legend cell align=left,fill=white, fill opacity=0.6, draw opacity=1,text opacity=1},
    	grid style={dashed}]
        ]
    % \addplot[color=red, mark=square, very thick]
    % coordinates {
    % (0,9.088809002749297)(1,9.58511885530115)(2,10.568747153421656)(3,10.533616446520677)(4,10.800969870254049)
    % };
    % \addlegendentry{\footnotesize{CNN+MLP}};
    
    \addplot[color=red, mark=o,,  very thick]   
    coordinates {
    (1000,0.69)  (10000,1.69)  (100000,2.86) (1000000,2.78) (10000000,2.82)  (100000000,2.75)

    };
    \addlegendentry{\footnotesize{MLP}};

    \addplot[color=blue, mark=square,  very thick]   
    coordinates {
    (1000,-0.38)  (10000,0.91)  (100000,2.11) (1000000,4.18) (10000000,4.36)   (100000000,4.38)

    };
    \addlegendentry{\footnotesize{biLSTM}};
    
        \addplot[color=green, mark=triangle,  very thick]   
    coordinates {
    (1000,-0.12)  (10000,0.23)  (100000,0.46) (1000000,0.28) (10000000,0.22)  (100000000,0.20)

    };
    \addlegendentry{\footnotesize{ESN}};
    
            \addplot[color=orange, mark=star,  very thick]   
    coordinates {
    (1000,0.1)  (10000,1.3)  (100000,3.03) (1000000,4.25) (10000000,4.41) (100000000,4.38)

    };
    \addlegendentry{\footnotesize{CNN+biLSTM}};

            \addplot[color=yellow,mark=diamond,  very thick]   
    coordinates {
    (1000,0.14)  (10000,1.61)  (100000,3.02) (1000000,3.81) (10000000,3.91) (100000000,4)  

    };
    \addlegendentry{\footnotesize{CNN+MLP}};
    \end{axis}
        \node[text width=2cm] at (1.1,1.9) 
    {T.1};
            \node[text width=2cm] at (2.1,2.8) 
    {T.2};
            \node[text width=2cm] at (3.6,2.55071082) 
    {T.3};
            \node[text width=2cm] at (5.0,5.35) 
    {T.4};
            \node[text width=2cm] at (6.4,5.45) 
    {T.5};
                \node[text width=2cm] at (7.4,5.45) 
    {T.6};
                \node[text width=2cm] at (1.65,0.2) 
    {DBP Region};
\draw[pattern=north west lines, pattern color=blue] (0,0) rectangle (0.5, 4.95);
    \end{tikzpicture}
    \caption{Simulation results}
    \label{numerical_result2} 
  \end{subfigure}%%
  \begin{subfigure}[b]{\linewidth/2}
    \centering
 \begin{tikzpicture}[scale=0.9]
    \begin{axis} [ylabel={Q-factor Gain [dB]}, 
        xlabel={\#Real Multiplications per Symbol},
        grid=both,
        xmode=log, 
        xmin=1e3,
        xmax=1e8,
    	    ymajorgrids = true,
    grid style = dashed,
    % 	xtick={1, ..., 20},
    % 	ymin=0.00001, ymax=0.1,
    %   ymode=log,
    %   log ticks with fixed point,
        legend style={legend pos=north west, legend cell align=left,fill=white, fill opacity=0.6, draw opacity=1,text opacity=1},
    	grid style={dashed}]
        ]
    % \addplot[color=red, mark=square, very thick]
    % coordinates {
    % (0,9.088809002749297)(1,9.58511885530115)(2,10.568747153421656)(3,10.533616446520677)(4,10.800969870254049)
    % };
    % \addlegendentry{\footnotesize{CNN+MLP}};
    
    \addplot[color=red, mark=o,,  very thick]   
    coordinates {
    (1000,0.42)  (10000,0.93)  (100000,1.92) (1000000,1.75) (10000000,1.71)  (100000000,1.69)

    };
    \addlegendentry{\footnotesize{MLP}};

    \addplot[color=blue, mark=square,  very thick]   
    coordinates {
    (1000,0.13)  (10000,0.27)  (100000,0.78) (1000000,1.36) (10000000,2.70)   (100000000,2.76)

    };
    \addlegendentry{\footnotesize{biLSTM}};
    
        \addplot[color=green, mark=triangle,  very thick]   
    coordinates {
 (1000,-0.12)  (10000,0.23)  (100000,0.46) (1000000,0.28) (10000000,0.22)  (100000000,0.20)

    };
    \addlegendentry{\footnotesize{ESN}};
    
            \addplot[color=orange, mark=star,  very thick]   
    coordinates {
    (1000,0.16)  (10000,0.49)  (100000,1.14) (1000000,1.85) (10000000,2.85) (100000000,2.80)

    };
    \addlegendentry{\footnotesize{CNN+biLSTM}};

            \addplot[color=yellow,mark=diamond,  very thick]   
    coordinates {
    (1000,0.3)  (10000,0.48)  (100000,1.06) (1000000,1.82) (10000000,2.2) (100000000,2.07)  

    };
    \addlegendentry{\footnotesize{CNN+MLP}};
    \end{axis}
        \node[text width=2cm] at (1.1,1.9) 
    {T.1};
            \node[text width=2cm] at (2.1,2.45) 
    {T.2};
            \node[text width=2cm] at (3.6,3.071082) 
    {T.3};
            \node[text width=2cm] at (5.0,4.1) 
    {T.4};
            \node[text width=2cm] at (6.4,5.45) 
    {T.5};
                \node[text width=2cm] at (7.4,5.45) 
    {T.6};
            \node[text width=2cm] at (1.65,0.2) 
    {DBP Region};
\draw[pattern=north west lines, pattern color=blue] (0,0) rectangle (0.5, 2.5);

    \end{tikzpicture}
    \caption{Experimental results}
    \label{experiment_result2} 
  \end{subfigure} 
  
  \caption{Q-factor gain dependence on the constrained multiplications number for the  equalizers having different architectures, presented in Sec.~\ref{sec:zoo}, in the case of DP-16 QAM single channel TWC-fiber 9$\times$50km. The power level is 2 dBm, which guarantees the high enough nonlinearity transmission regime.}\label{fig:Result_02}
\end{figure*}

When considering how NN equalizers function with the experimental data, Fig.~\ref{experiment_result1}, we can mention two major observations. First, similarly to the numerical results, the CNN+biLSTM is best-performing among all the considered NN structures in terms of the Q-factor gain. The CNN+biLSTM demonstrated a 2.91 dB improvement when compared to the conventional DSP, 0.15~dB when compared to the biLSTM equalizer, 0.61~dB when compared to the CNN+MLP equalizer, 0.96~dB when compared to the MLP equalizer, and 2.33 dB when compared to the ESN equalizer. Additionally, as was also observed in the numerical analysis, a lower number of training epochs was necessary to reach the best performance point when we add a convolutional layer: using the CNN+biLSTM we needed 169 epochs, while for the pure biLSTM this number was 232 epochs; the number of epochs required for the CNN+MLP to reach the best performance was 107, and for the pure MLP it was 753 epochs. Second, compared to the simulation, the overall gain of all NN-based equalizers is slightly reduced. This can be explained by the existing ``reality gap'' between the numerical model and the true experimental transmission results. In real transmission, extra nonlinearity and non-ideally behavior of the transceivers (signal clipping by the ADC/DAC, harmonic and intermodulation distortions of the driver amplifier (DA), I/Q skew, etc.) add extra noise and complexity to the process of channel inversion. We believe that with just the split-step method, the NNs can unroll the synthetic propagation effect more easily than reverting the actual propagation in the experimental condition. We also point out that even though the gain numbers are different in the numerical and experimental data, the NN structures' performance followed the same pattern for both numerical and experimental data: the best performance was attributed to the CNN+biLSTM, the next level performance pertains to the biLSTM, followed by the CNN+MLP, the MLP and, finally, the ESN. 

Finally, of all equalizer types investigated in this study, the DBP 3 StPS applied with two samples per symbol was still the least complex method. In all simulation and experiment test cases, however, the CNN+biLSTM outperformed the DBP, as shown in Fig.~\ref{Result_01}.  Even by optimizing the DBP's nonlinear coefficient parameter ($\gamma$), the DBP approach was able to enhance the Q-factor by 1.32 dB, whereas the CNN+biLSTM equalizer improved by 2.91 dB, in the experimental case. The boost in performance by the CNN+biLSTM relative to the DBP in the experiment scenario demonstrated the NN-equalizer power in mitigating transmission impairments in a practical application.

\subsection{Comparative analysis of different NN-based equalizers  with the fixed computational complexity}

The analysis given above does not address the question of which NN topology would provide the best gain if we restrict the NN structure's complexity to a certain level. To answer this question, we retested the equalizers constraining the total number of real multiplications per recovered symbol (RMpS). We considered the complexity values in the range from $10^{3}$ to $10^{8}$ RMpS. We note that the NN structures with large RMpS ($\sim 10^{8}$) can be prohibitively complex for efficient hardware implementation. However, Ref.~\cite{que2020mapping} demonstrated an efficient FPGA implementation of LSTMs with 256 and 512 hidden units. This result reveals that the architectures outlined in this research are still feasible for realistic signal processing when advanced techniques for NN hardware implementation are used.

The hyper-parameters distributions for each NN architecture with the complexity constraint are summarized in Table~\ref{table_hyperparameters} in the cells marked from '`Topology 1'' to ``Topology 6''. The parameters of those topologies were also tuned by the BO: for each case, we reduced the allowed BO search range to comply with each computational complexity constraint.

As seen in Fig.~\ref{fig:Result_02}, for different allowed computational complexity levels, the performance ranking of equalizer types changes. Several conclusions can be drawn analyzing the results emerging from the simulated (Fig. \ref{numerical_result2}) and experimental (Fig. \ref{experiment_result2}) data. First, in the experimental scenario, the best complexities corresponding to the maximum gain coincide with the complexities identified by the BO procedure, which confirms the effectiveness of the BO in finding the ``right'' NN architecture. Second, in simulations, the maximum performance is reached already at a lower complexity level compared to the experimental results. As it can be seen from the experimental figure, the CNN+biLSTM, CNN+MLP, and biLSTM equalizers need $\approx 10^7$ RMpS, while in the simulation $\approx 10^6$ RMpS was already enough to achieve the best performance. This observation further confirms that the NN can cope with the reversion of the simulated channel more easily than with the reversion of experimentally obtained data. Third, when we increase the complexity above the level determined by the BO, the gain remains nearly constant: this is due to overfitting and it is particularly pronounced in the MLP scenario. The key concept of the function approximation capability of the MLP belongs to its number of i) feed-forward hidden layers and ii) hidden neurons; these two parameters define the NN's capacity \cite{Goodfellow-et-al-2016}. Changing, the MLP's capacity by adjusting the complexity levels frequently leads to unpredictable changes in the NN's performance. Starting at the $10^5$ complexity level for both simulation and experimental layouts, we can see that MLPs with oversized capacity suffer from overfitting, as the network memorizes the properties of the training set in such detail that it can no longer efficiently recover the information from the inference dataset~\cite{Goodfellow-et-al-2016}. The latter blockades the equalizer from providing further Q-factor improvement. Thus, we argue that the architectures found by the BO identify the most appropriate NN equalizer's capacity (structure) matching our problem, and a further increase in complexity cannot render any noticeable performance improvement. 

Next, we note that for the high level of RMpS (Topologies 4, 5, and 6), the best-performing equalizer is the CNN+biLSTM. However, once we reduce the number of real multiplications from Topology 3 and below, the best-performing equalizer turns out to be the traditional MLP. This can be explained by the fact that advanced architectures, such as CNN and biLSTM, require more filters and a higher number of hidden units, respectively, to learn the complete dynamics of the data. Also, we observe that the CNN+biLSTM performs similarly to the CNN+MLP at low complexity levels (orange and yellow curves in Fig.~\ref{fig:Result_02}), and similarly to the biLSTM (blue line) at high complexity. Consequently, we can infer how the addition of a convolutional layer works: while for high complexity the blue and orange curves are approximately the same, at a lower allowed complexity level the CNN+biLSTM performs better.

In addition, we used the hatched blue zone in both simulation and experimental cases, for the traditional DBP with 3 StPS, to highlight the performance of the NN equalizers with similar CC of the DBP. Then it is evident that reducing the number of neurons, filters, and hidden units is not the optimal technique to achieve low complexity architectures, because performance fell below the DBP level. As a possible alternative,  pruning and quantization techniques \cite{cheng2017survey,Long_2019} can be used to minimize the CC of the NN equalizers without compromising their performance, making the NN equalizers appealing not only for their good performance but also for their decreased complexity.

Finally, the performance shown by the ESN does not meet the expectations, showing the lowest achievable gain number. However,~\cite{brain-inspired} contains the results explaining the poor ESN performance for the nonlinear wireless scenario. It was shown that in the channel with a high level of noise, the ESN-based equalizer was found to perform poorly. Furthermore, they demonstrated that by increasing the ESNs' number of neurons (i.e., the complexity), and, thus, effectively increasing the hidden dimensionality of the representation, equalization performance worsens. Moving to the nonlinear optical channel equalization, we observed both aforementioned effects: the performance was relatively poor due to the high level of noise, and the performance did not improve when we increased the complexity, as observable in the green curve of Fig.~\ref{fig:Result_02}. 

\section{Conclusion}

In this paper, we proposed and examined novel designs of combined NNs: (a) CNN+MLP and (b) CNN+biLSTM for the equalization of coherent optical fiber channels. We reviewed and compared several key existing NN-based methods with the proposed new algorithms using both the numerically simulated synthetic data and the experimental data from the benchmark transmission system. One of the most relevant outcomes of our work lies in the reported analytical expressions for the complexity (the number of real multiplications) associated with each NN type considered in the paper. 
Although a comparative analysis has been carried out for a specific benchmark system, we believe that our findings are relatively generic and can be applied to other scenarios.

% The experimental benchmark system for comparison of different channel equalizers was designed to operate in a nonlinear regime, i.e., fiber Kerr nonlinearity was the main source of signal degradation. More specifically, we considered a low dispersion TWC fiber, and we processed the data at 2 dBm signal launch power to induce clear nonlinear signal distortions. We stress that the conclusions of the trade-off for each NN equalizer performance and complexity are specific for the particular considered system in this paper. However, we believe that our research paves the way for a methodology for calculating the computational cost of different NN-based channel equalizers. 

Fiber Kerr nonlinearity was the predominant source of signal deterioration in the experimental benchmark system for comparing different channel equalizers. In order to produce clear nonlinear signal distortions, we used a low dispersion TWC fiber and processed the data at 2 dBm signal launch power. We emphasize that the trade-off conclusions for each NN equalizer's performance and complexity are unique to the system under consideration in this paper. However, we believe that our research paves the way for a methodology for estimating the computational cost of various NN-based channel equalizers.

% First, we described in detail the design of the selected most promising NN-based equalizers. We note that, recently, the architectures using recurrent NN have become especially popular due to their capability of handling the channel memory. In our work, we introduced the new composite NN equalizer designs that combine a convolutional layer with either densely connected layers (CNN+MLP) or with a biLSTM layer (the CNN+biLSTM). To derive the best-performing structures, we utilized the Bayesian optimization of each NN type that provides the optimized set of hyper-parameters for each particular NN structure. For these optimized structures, we found that the best performance of the test system was rendered by the new CNN+biLSTM architecture, though the performance of the pure biLSTM was only slightly lower. However, the optimized CNN+biLSTM design corresponded to the highest complexity among all cases studied. 

We described in detail the design of the selected most promising NN-based equalizers. To derive the best-performing NN structures, we utilized the Bayesian optimization of each NN type that provides the optimized set of hyper-parameters for each particular NN-based equalizer. For these optimized structures, we found that the best performance of the test system was rendered by the new CNN+biLSTM architecture, though the performance of the pure biLSTM was only slightly lower. However, the optimized CNN+biLSTM design corresponded to the highest complexity among all cases studied. 

The important part of the analysis was the comparison of the performance under the condition of the restricted complexity: the respective results are given in the last section. We found that at high complexity levels the best-performing NN among studied cases is the CNN+biLSTM. However, when reducing the complexity, we observed the transition: when the allowed complexity is relatively low, the best-performing structure turned out to be the simple MLP. We can explain this behavior as follows: the advanced architectures (the CNN and biLSTM) require more complexity-hungry components (filters or hidden units) to learn the data dynamics, while the MLP is less demanding using just the summation and activation functions at the basic level. Overall, we conclude that the addition of the convolutional layer can be beneficial if we do not restrain the complexity. However, the important message is that complexity can play an important and even crucial role in the hardware implementation of the NN equalizers. Our analysis demonstrates that even the simple NN structures, like the MLP, can outperform the more advanced counterparts when the complexity is constrained to relatively low levels.

\bibliographystyle{IEEEtran}
\bibliography{references}

\end{document}